\documentclass[reprint,aps,prd,nofootinbib,twocolumn,superscriptaddress,preprintnumbers]{revtex4-2}

\usepackage[T1]{fontenc}
\usepackage[usenames,dvipsnames]{xcolor}
\usepackage{aas_macros,graphicx,hyperref,mathrsfs,orcidlink,subfigure}
\hypersetup{colorlinks=true,citecolor=red,linkcolor=red,urlcolor=blue}

\usepackage{ulem}

\usepackage{graphicx,amssymb,amsmath,amsthm,amsfonts,epstopdf,epsfig,epsf,times}

\usepackage{amsmath}

\usepackage{esint}

\newcommand{\pa}[1]{\left(#1\right)}

\newcommand{\be}{\begin{equation}}
\newcommand{\ee}{\end{equation}}
\newcommand{\bea}{\setlength\arraycolsep{2pt} \begin{eqnarray}}
\newcommand{\eea}{\end{eqnarray}}

\newcommand{\md}{\mathrm{d}}

\def\be{\begin{equation}}
\def\ee{\end{equation}}
\def\bag{\begin{aligned}}
\def\eag{\end{aligned}}
\def\bea{\begin{eqnarray}}
\def\eea{\end{eqnarray}}
\def\ba{\begin{array}}
\def\ea{\end{array}}

\def\bc{\begin{center}}
\def\ec{\end{center}}

\begin{document}

\title{Autocorrelation in black hole flare movies: Departures from Kerr}

\author{Fang-Stars Wei}
\affiliation{Institute of Fundamental Physics and Quantum Technology, and School of Physical Science and Technology, \\Ningbo University, Ningbo, Zhejiang 315211, China}

\author{Zelin Zhang}
\affiliation{Institute of Fundamental Physics and Quantum Technology, and School of Physical Science and Technology, \\Ningbo University, Ningbo, Zhejiang 315211, China}
\affiliation{Zhejiang Key Laboratory of Extreme Universe, and BINGO Center, \\Ningbo University, Ningbo, Zhejiang 315211, China}

\author{Zhenyu Zhang}
\email{Contact author: zhangzhenyu@nbu.edu.cn}
\affiliation{Institute of Fundamental Physics and Quantum Technology, and School of Physical Science and Technology, \\Ningbo University, Ningbo, Zhejiang 315211, China}
\affiliation{Zhejiang Key Laboratory of Extreme Universe, and BINGO Center, \\Ningbo University, Ningbo, Zhejiang 315211, China}

\author{Bin Chen}
\email{Contact author: chenbin1@nbu.edu.cn}
\affiliation{Institute of Fundamental Physics and Quantum Technology, and School of Physical Science and Technology, \\Ningbo University, Ningbo, Zhejiang 315211, China}
\affiliation{Zhejiang Key Laboratory of Extreme Universe, and BINGO Center, \\Ningbo University, Ningbo, Zhejiang 315211, China}
\affiliation{School of Physics, and Center for High Energy Physics, Peking University, No.5 Yiheyuan Rd, Beijing 100871, China}

\begin{abstract}
Autocorrelation in black hole movies encodes the time delays and angular separations between gravitationally lensed images, offering a means of testing the spacetime geometry. Using an orbiting hot spot in the parametrized Johannsen-Psaltis metric with a single deformation parameter, we investigate the ability of lensing-induced correlations to distinguish departures from Kerr. 
From ray-traced movies, we extract the slope (k) and intercept (b) of an approximately linear secondary ridge in the time-angle autocorrelation map as potential future observables. Variations in the spin and the metric deformation parameter produce distinctive trajectories in the $(k,b)$ plane, allowing the two observables to jointly constrain both parameters when the orbital radius, observer inclination, and orbital angular velocity are specified. By varying the orbital angular velocity, we construct a Kerr reference region in the $(k,b)$ plane, and we find that, for some low-spin configurations with negative deformation parameters, the measured slope-intercept pairs of the secondary peaks lie outside this reference region. This feature appears for sub-Keplerian, Keplerian, and super-Keplerian motion, indicating that secondary-ridge measurements from black hole flare movies have the potential to distinguish non-Kerr signatures.
\end{abstract}

\maketitle

\section{Introduction}
\label{sec:introduction}

The Event Horizon Telescope (EHT) images of M87* and Sgr A* \cite{EventHorizonTelescope:2019dse, EventHorizonTelescope:2022wkp} have opened new avenues for studying black hole physics on horizon scales \cite{Vagnozzi:2022moj,Chen:2022scf,He:2024amh,Zhang:2024jrw,Zhang:2026zyq}. In the near future, horizon-scale black hole observations are expected to move from static, time-averaged images toward time-resolved measurements, yielding black hole movies \cite{johnson2017dynamical,EventHorizonTelescope:2024whi,Ayzenberg:2023hfw}. By adding a temporal dimension, these movies would substantially expand the information available from black hole images. An important question is how to extract physical information from these future movies, particularly how to disentangle gravitational effects from those of the astrophysical environment \cite{Johnson:2019ljv,Himwich:2020msm,Gralla:2020srx,Lara:2021zth,Kocherlakota:2022jnz,Paugnat:2022qzy,Palumbo:2024czv,Hou:2024qqo}. In this context, autocorrelation statistics offer a promising approach.

In the one-dimensional autocorrelation of a light curve, gravitational lensing can give rise to secondary peaks associated with light echoes \citep{Fukumura:2007xr,Moriyama:2019mhz,Chesler:2020gtw,Wong:2024gph}. These echoes arise because radiation from the same emission event reaches the observer along multiple paths with different travel times \cite{Wong:2020ziu}. However, such secondary peaks have not been identified in the observed millimeter light-curve autocorrelation of Sgr A* \cite{EventHorizonTelescope:2022wkp}. This is because the lensing-induced correlations are exponentially suppressed at a rate governed by the Lyapunov exponent and can be easily 
obscured by the primary peak when the correlation timescale of internal astrophysical processes exceeds the characteristic lensing time delay \cite{Cardenas-Avendano:2024sgy}.

Retaining spatial information offers a way to recover lensing signatures that are obscured in light-curve autocorrelations. For a Kerr black hole viewed near its rotation axis, analytical calculations predict a multi-peak structure in the plane of time and azimuthal lags \cite{Hadar:2020fda}. The shape of each peak reflects the statistical properties of the emission, whereas their positions and relative heights encode information about the background spacetime \cite{Gralla:2019drh}. 

The prospects for extracting lensing signatures through image correlations have also been investigated numerically. In \cite{Bezdekova:2025wii}, distinct lensing-induced correlation enhancements were identified in the two-point image correlation function of an M87* movie generated from a general relativistic magnetohydrodynamic simulation with slow-light ray tracing. These features remain visible even after the movie is blurred to an angular resolution of approximately $15\,\mu\mathrm{as}$, representative of the nominal resolution anticipated for the next-generation EHT.

Observational limitations impose requirements on monitoring duration and temporal cadence, but sufficiently bright, localized flares may allow lensing information to be extracted from shorter observations \cite{Bezdekova:2026zgq}. A complementary question is how much lensing information can be extracted from a single, sufficiently bright and localized flare. For such localized emission, the autocorrelation of black hole flare movies has been investigated using Keplerian hot-spot models. In \cite{Zhang:2025vyx}, two-dimensional autocorrelation maps in time and azimuthal lag were analyzed for a hot spot orbiting a Kerr black hole viewed near the rotation axis. The lensing-induced secondary peaks approximately intersect at a fixed point whose position depends on the black hole spin and is comparatively insensitive to the hot spot's orbital radius, suggesting a means of constraining spin within the adopted model. Cross-correlations involving images up to eighth order were also examined in \cite{Zhu:2025jqh}, revealing that higher-order correlation structures are less sensitive to the source's orbital configuration than their lower-order counterparts.

The studies discussed above establish autocorrelation in black hole movies as a probe of strong gravitational lensing in the Kerr geometry. A natural next step is to determine whether these correlations can distinguish departures from Kerr and thereby contribute to tests of gravity in the strong-field regime. Here we investigate this question using the Johannsen-Psaltis spacetime \cite{Johannsen:2011dh}, with a single nonzero deformation parameter $\epsilon_3$ as a test case \cite{Li:2014fza,Wang:2025ihg}. We generate movies of a compact hot spot on an equatorial circular orbit and characterize the lensing-induced secondary ridge by its slope and intercept in the time-angle autocorrelation map. We first show that, with the orbital radius, observer inclination, and ratio of orbital to Keplerian angular velocity specified, measurements of the slope and intercept can jointly constrain spin and deformation. To test whether non-Kerr signatures remain distinguishable when the assumption of Keplerian motion is relaxed \cite{Matsumoto:2020wul,Porth:2020txf,Lin:2023tog}, we construct a Kerr reference region by varying both the spin and the orbital angular velocity at fixed orbital radius and observer inclination. We find that some low-spin configurations with negative $\epsilon_3$ produce slope-intercept pairs outside the Kerr reference region, not only for Keplerian but also for sub-Keplerian and super-Keplerian orbital motions.

The remainder of the paper is organized as follows. Section~\ref{sec:setup} introduces the autocorrelation statistic and describes the spacetime, emission model, and numerical method. Section~III presents the secondary-ridge observables, their parameter dependence, and the comparison with Kerr for non-Keplerian source motion. Section~IV summarizes the results and their implications for probing departures from Kerr. Throughout the paper, we use geometrized units $G=c=1$ and measure lengths and times in units of the black hole mass $M$.

\section{Setup and numerical method}
\label{sec:setup}

\subsection{Autocorrelation of black hole movies}
\label{subsec:autocorrelation}

The time-resolved black hole image can be characterized by the intensity $I(t,\rho,\varphi)$, where $t$ is the observational time and  $(\rho,\varphi)$ are polar coordinates on the screen. For a statistically stationary and axisymmetric system, the two-point correlation function depends only on the time lag $T=t'-t$ and the angular displacement $\Phi=\varphi'-\varphi$.  The autocorrelation is therefore given by
\bea
\begin{aligned}
 C(T,\Phi)=&\\
 \int \rho\,\md\rho&\int \rho'\,\md\rho'
 \left\langle
 \Delta I(t,\rho,\varphi)
 \Delta I(t+T,\rho',\varphi+\Phi)
 \right\rangle ,
 \label{eq:autocorrelation}
\end{aligned}
\eea
where the intensity fluctuation is defined as
\begin{align}
 \Delta I(t,\rho,\varphi)
 =I(t,\rho,\varphi)
 -\left\langle I(t,\rho,\varphi)\right\rangle ,
 \label{eq:intensity_fluctuation}
\end{align}
and
\begin{align}
 \left\langle \right\rangle
 =\lim_{\tau\rightarrow\infty}\frac{1}{2\pi\tau}
 \int_0^\tau\md t\int_0^{2\pi}\md\varphi\, .
 \label{eq:ensemble_average}
\end{align}
The two-dimensional autocorrelation exhibits a reflection symmetry, i.e., $C_{2D}\left(T,\Phi\right) = C_{2D}\left(-T,-\Phi\right)$. Its maximum is attained when there is no separation in either time or angle, namely, $C_{2D,\text{max}}\pa{T,\Phi}=C_{2D}\pa{0,0}$.

For numerical calculations, the continuous integrals above are replaced by discrete Riemann sums. Suppose that adjacent frames in the black hole movie are separated by a time interval $t_0$, and that the coordinates $\pa{t,\rho,\varphi}$ are sampled with $\pa{n_t,n_{\rho},n_{\varphi}}$ discrete elements, respectively. The corresponding discretized form of the autocorrelation is then
\bea\label{eq:numAC} 
\begin{aligned}
	C_{2D}&\pa{T,\Phi}\simeq N\sum_{i=1}^{n_{\rho}}\sum_{i^{\prime}=1}^{n_{\rho}}\sum_{j=1}^{n_{\varphi}}\sum_{k=1}^{n_{t}}\frac{i}{n_{\rho}}\frac{i^{\prime}}{n_{\rho}}\frac{1}{n_\varphi}\frac{1}{n_t}\\ 
	&\Delta I\left(k t_0,\frac{i}{n_{\rho}},\frac{2\pi j}{n_{\varphi}}\right)\Delta I\left(kt_0+T,\frac{i^{\prime}}{n_{\rho}},\frac{2\pi j}{n_{\varphi}}+\Phi\right)\,, 
\end{aligned}
\eea 
where $N$ denotes an overall normalization factor. We further introduce the normalized autocorrelation through $C\left(T,\Phi\right) \equiv C_{2D}\pa{T,\Phi}/C_{2D}\pa{0,0}$ to remove the factor $N$.

\subsection{Johannsen-Psaltis spacetime}

The Johannsen-Psaltis (JP) spacetime provides a model-independent parametrization of possible deviations from the Kerr geometry. We adopt it here to investigate their observational signatures. In Boyer-Lindquist coordinates, the JP metric can be written as
\bea
\begin{aligned}
\md s^2=&-H(r,\theta)\left(1-{2Mr \over \Sigma}\right)\md t^2
+{\Sigma H(r,\theta) \over \Delta+h(r,\theta)a^2\sin^2\theta} \md r^2\\
& +\Sigma \md \theta^2 -{4aMr\sin^2\theta \over \Sigma}H(r,\theta) \md t \md \phi \\
& +\sin^2\theta \left[ \Sigma+H(r,\theta){a^2 \left(\Sigma+2Mr\right)\sin^2\theta \over \Sigma} \right] \md \phi^2 ,
\end{aligned}
\label{eq:JP_metric}
\eea
where
\begin{align}
	\Sigma=r^2+a^2\cos^2\theta, \qquad
	\Delta=r^2-2Mr+a^2 .
\end{align}
Here $M$ and $a$ represent the mass and spin of the JP black hole, respectively. The function $H(r,\theta)$ is defined as
\begin{align}
	H(r,\theta)=1+h(r,\theta),
	\label{eq:JP_H}
\end{align}
where
\begin{align}
	h(r,\theta) \equiv \sum_{i=0}^{\infty}
	\left(\epsilon_{2i}+\epsilon_{2i+1}{Mr \over \Sigma}\right)
	\left({M^2 \over \Sigma}\right)^i .
	\label{eq:JP_h_general}
\end{align}
If we require the metric to asymptotically approach the Schwarzschild spacetime in the far-field region, $\epsilon_0$ and $\epsilon_1$ must vanish. Furthermore, the Lunar Laser Ranging experiment constrains $\epsilon_2$ to be very close to zero ($\lesssim 0.00046$) \cite{Williams:2004qba}. Therefore, in this work, we consider only the next-highest-order deformation parameter $\epsilon_3$. The function $h(r,\theta)$ then reduces to
\begin{align}
	h(r,\theta)=\epsilon_3 {M^3 r \over \Sigma^2}.
	\label{eq:JP_h}
\end{align}
When $\epsilon_3=0$, the metric reduces to the Kerr metric, while for $a=0$, the spacetime becomes spherically symmetric. Only in these two limiting cases can the null geodesic equations be separated, allowing them to be solved analytically. In the general case, we must resort to numerical ray tracing.

\begin{figure*}[t]
  \centering
  \includegraphics[width=\textwidth]{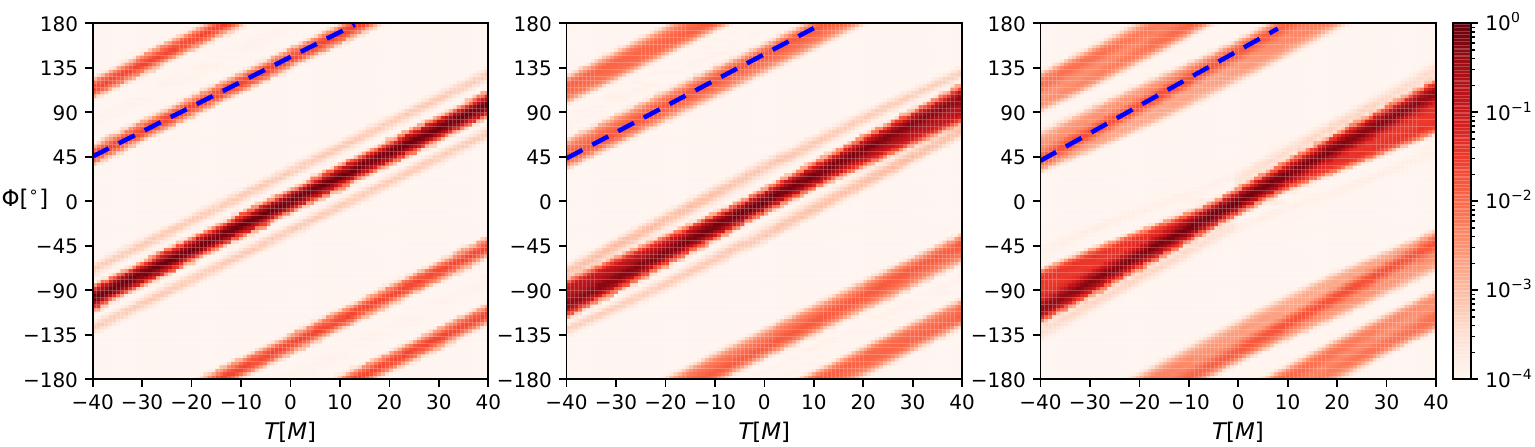}
\caption{Autocorrelation results for $a=0.94$, $r_s=8M$, and $\epsilon_3=-5$. In each panel, the blue dashed line marks the fit to the $0<b<180^\circ$ branch of the secondary peaks. Left: $\theta_o=0.1^\circ$. Middle: $\theta_o=17^\circ$. Right: $\theta_o=30^\circ$.}
\label{fig:cora094r8}
\end{figure*}

The location of the horizon \footnote{Following the terminology of \cite{Johannsen:2013rqa}, we refer to this surface as the Killing horizon and use it as the ray-tracing termination boundary.} is determined by the equation
\begin{align}
	g_{t\phi}^2-g_{tt}g_{\phi\phi}=0,
	\label{eq:JP_horizon_condition}
\end{align}
which can be written as
\begin{align}
H(r,\theta)\left[\Delta+h(r,\theta)a^2\sin^2\theta\right]=0.
	\label{eq:JP_horizon}
\end{align}
However, for certain values of $\epsilon_3$ and $a$, this equation admits no real solution, indicating that the JP spacetime does not necessarily possess a horizon and may instead describe a naked singularity. In this work, we restrict our analysis to the parameter space in which Eq.~\eqref{eq:JP_horizon} admits real roots.

\subsection{Orbiting hot spot and dynamic imaging}
\label{subsec:hotspot_movie}
We model the flare source as a compact hot spot moving on a circular orbit in the equatorial plane of the black hole \cite{Broderick:2005jj,Liu:2014awa,gravity2018detection,Yfantis:2025jtw,Cruz-Osorio:2026osq,Yfantis:2026mnu}. For a general stationary and axisymmetric metric, the angular velocity of equatorial circular geodesics is
\begin{align}
	\omega_\pm =
	{-\partial_r g_{t\phi}
	\pm
	\sqrt{(\partial_r g_{t\phi})^2
	-\partial_r g_{tt}\partial_r g_{\phi\phi}}
	\over
	\partial_r g_{\phi\phi}}
	\bigg|_{r=r_s,\theta=\pi/2}.
	\label{eq:Omega_circular}
\end{align}
The plus and minus signs correspond to prograde and retrograde circular orbits, respectively. In this work, we focus on prograde orbits and denote the corresponding angular velocity by $\omega_K=\omega_+$.

We further assume that the emission is concentrated around the hot-spot center, 
with the emissivity profile \cite{GRAVITY:2020lpa}
\begin{align}
	J\propto\exp\left(-\frac{x^2}{2s^2}\right),
	\label{eq:hotspot_emissivity}
\end{align}
where $x$ denotes the spatial separation between an emitting point and the 
instantaneous center of the hot spot. Because the hot spot moves along its 
orbit, $x$ depends on both the position and the emission time, and hence $J$ 
describes a time-dependent emission pattern. The parameter $s$ controls the 
physical size of the hot spot, and we set $s=0.3M$ in our numerical 
calculations. Assuming optically thin emission and negligible absorption, the 
observed intensity is computed by integrating the redshifted emissivity along 
each null ray,
\begin{align}
	I_o\propto\int g^2J\,\mathrm d\lambda,
	\label{eq:observed_intensity}
\end{align}
where $g=\nu_o/\nu_e$ is the redshift factor and $\lambda$ is an affine 
parameter. The details of the GRRT prescription are provided in 
Appendix~\ref{app:hotspot_emission}.

\section{Secondary ridge and measured observables}

Figure~\ref{fig:cora094r8} presents the numerical results of the autocorrelation for $a=0.94$, $r_s=8M$, and $\epsilon_3=-5$. Consistent with \cite{Zhang:2025vyx}, the autocorrelation results exhibit a multi-peak structure, with the peak values at different orders differing by roughly an order of magnitude. The maximum autocorrelation occurs at the central peak passing through the origin, with two secondary peaks symmetrically distributed on either side, whose values are approximately $0.01$--$0.06$ times the maximum value \footnote{The very weak peaks closest to the central peak correspond to higher-order structures and are not considered here.}. When the inclination is close to zero, the autocorrelation peaks of different orders are nearly straight and approximately parallel to one another. As the inclination increases, the peak shapes gradually distort and exhibit a ridge-like form. Therefore, extracting the approximate slope and intercept of the secondary peak from these results requires a weighted least-squares fit, with the autocorrelation values serving as the weights.

Specifically, we select a region $\mathcal{R}$ containing one continuous secondary branch while excluding all other peaks, and approximate the center line of this branch by
\begin{align}
\Phi=kT+b,
\label{eq}
\end{align}
where $k$ and $b$ denote the measured slope and intercept, respectively. These parameters are determined by minimizing
\begin{align}
\chi^2(k,b)
=
\sum_{(T_i,\Phi_i)\in\mathcal{R}}
C(T_i,\Phi_i)\left[\Phi_i-kT_i-b\right]^2,
\end{align}
which yields
\begin{align}
k&=
\frac{\langle T\Phi\rangle_C-\langle T\rangle_C\langle\Phi\rangle_C}
{\langle T^2\rangle_C-\langle T\rangle_C^2},
\qquad
b=\langle\Phi\rangle_C-k\langle T\rangle_C,
\end{align}
where the autocorrelation-weighted average is defined as
\begin{align}
\langle X\rangle_C
\equiv
\frac{
\sum_{(T_i,\Phi_i)\in\mathcal{R}}C(T_i,\Phi_i)X_i
}{
\sum_{(T_i,\Phi_i)\in\mathcal{R}}C(T_i,\Phi_i)
}.
\end{align}
Since there are two secondary branches, we choose the branch with $0<b<180^\circ$.

In the face-on limit, the slope satisfies $k=\omega_K$. For a finite inclination, the slope still retains part of the physical meaning of the angular velocity, corresponding to the ``pattern speed'' introduced in \cite{Conroy:2023kec,Conroy:2025nev}. In the following, we regard $k$ and $b$ as observable quantities that can in principle be extracted from future black hole flare movies, and can therefore be directly compared with theoretical predictions.

\subsection{One-parameter response in the slope-intercept plane}
\label{subsec:parameter_response}

From now on, we fix the observer inclination at $\theta_o=17^\circ$, which is representative of the viewing angle of M87*. For Sgr A*, the inclination remains uncertain. However, based on numerical simulations and observational analyses, the EHT Collaboration tends to favor a relatively low-inclination configuration ($\theta_o\lesssim 30^\circ$) \cite{EventHorizonTelescope:2022wkp}. Moreover, at low inclinations, the secondary ridges remain approximately linear, allowing them to be characterized by their slopes and intercepts.
Therefore, our choice provides a representative setting for our analysis.

\begin{figure}[thbp!]
	\centering
	\includegraphics[width=\columnwidth]{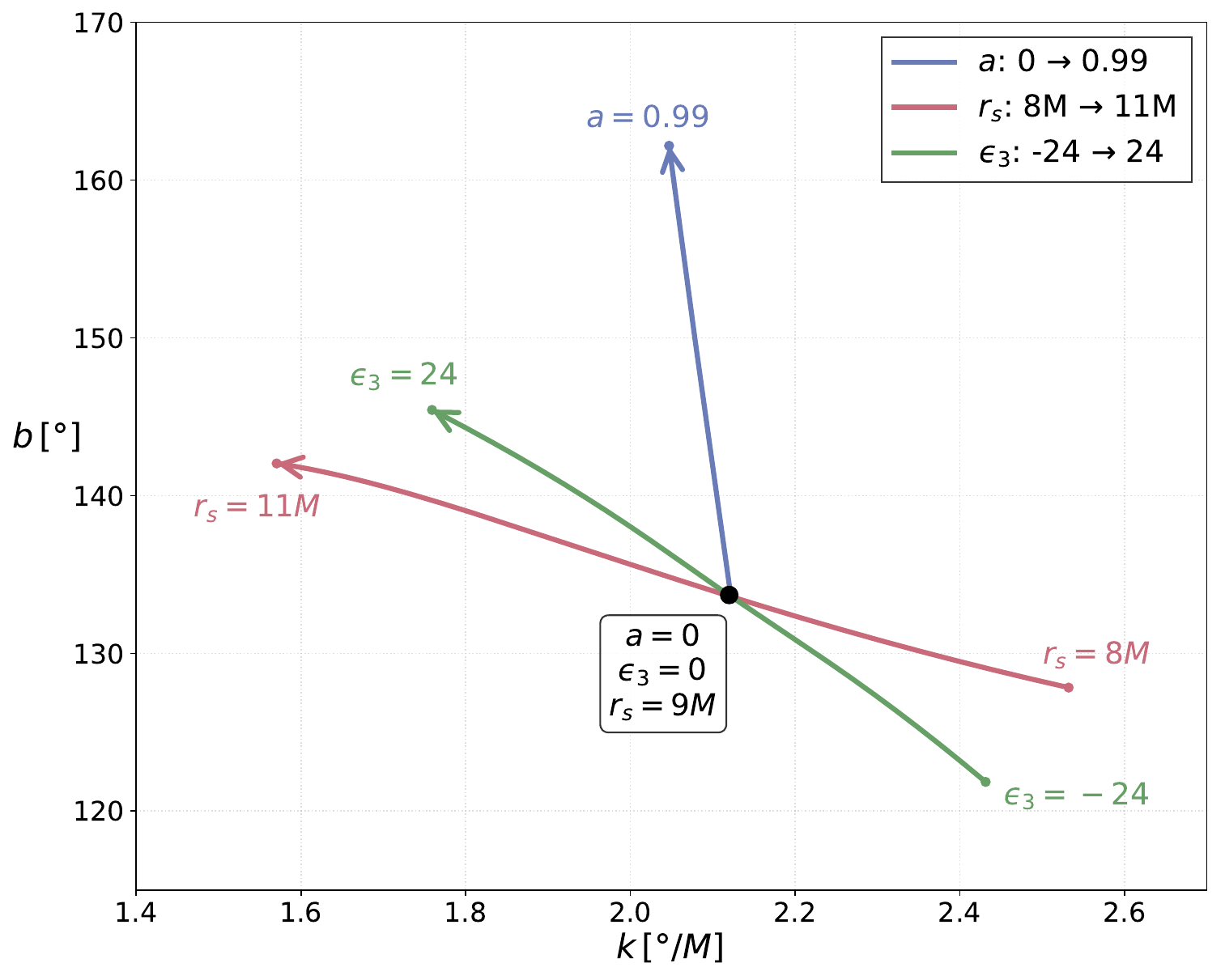}
	\caption{Response of the secondary-ridge slope and intercept to the spin $a$ (blue), hot-spot orbital radius $r_s$ (red), and JP deformation $\epsilon_3$ (green). The inclination is fixed at $\theta_o=17^\circ$.}
	\label{fig:parameter_flow}
\end{figure}

Fig.~\ref{fig:parameter_flow} compares the displacement of $(k,b)$ when the spin, source radius, or JP deformation is varied separately. The common reference model is $a=0$, $\epsilon_3=0$, and $r_s=9M$. The blue trajectory increases the spin from $0$ to $0.99$, the red trajectory increases the orbital radius from $8M$ to $11M$, and the green trajectory changes $\epsilon_3$ from $-24$ to $24$. 

At fixed $r_s$ and $\epsilon_3$, increasing the spin $a$ mainly increases the intercept while changing the slope comparatively little.

Increasing $r_s$ at fixed spin and deformation produces a pronounced decrease in slope, as expected from the lower orbital angular velocity, together with an increase in intercept. Since $r_s$ affects both $k$ and $b$, uncertainty in the orbital radius propagates directly into any inference of the metric.

Increasing $\epsilon_3$ shifts the prediction from the large-slope, small-intercept side toward the small-slope, large-intercept side. This trajectory is partly aligned with the radius response but is not identical to it, because $\epsilon_3$ modifies both timelike and null geodesics. The nonparallel components are precisely the information needed to distinguish a metric deformation from a change in the source orbit.

\begin{figure*}[t]
  \centering
  \includegraphics[width=\textwidth]{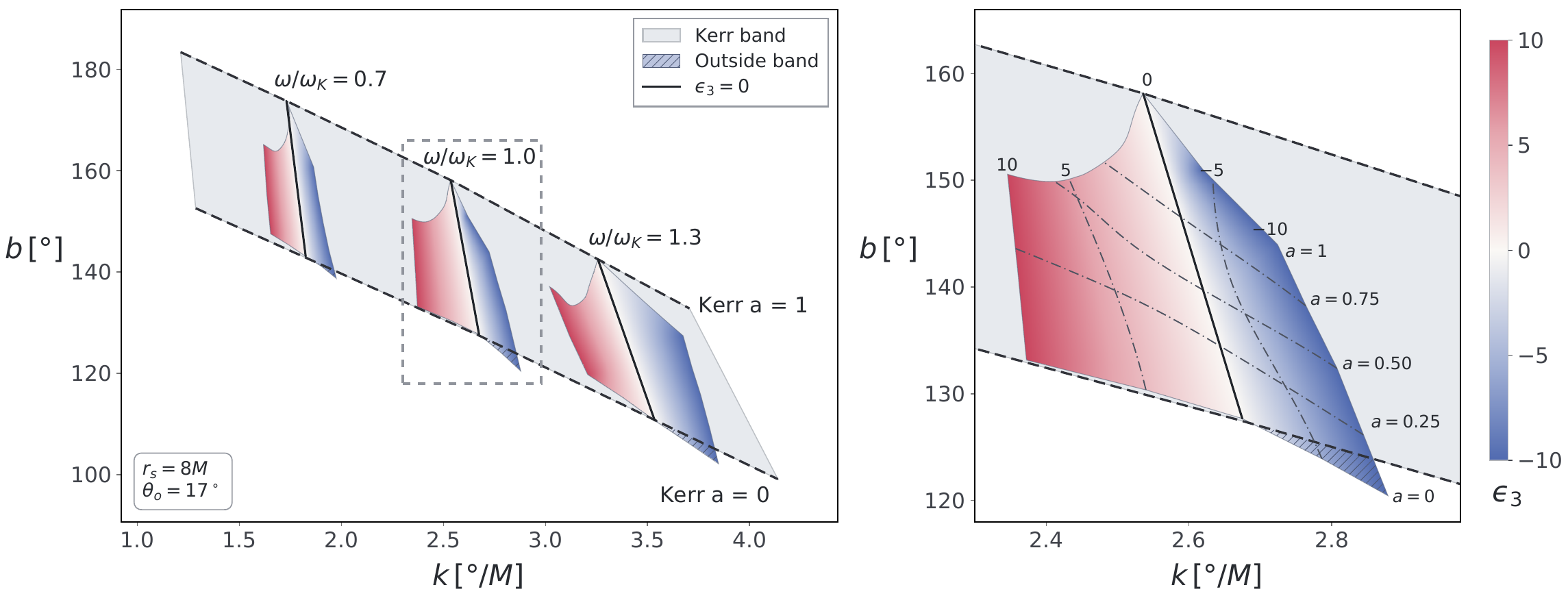}
\caption{
Secondary-ridge slope $k$ and intercept $b$ for $r_s=8M$ and $\theta_o=17^\circ$.
Left: the gray Kerr reference band lies between the dashed $a=1$ and $a=0$ curves, each obtained by varying $\omega/\omega_K$ from $0.5$ to $1.5$.
The colored regions show JP predictions for $\omega/\omega_K=0.7$, $1.0$, and $1.3$, with $\omega_K$ evaluated in the corresponding spacetime.
Color denotes $\epsilon_3\in[-10,10]$, while spin varies over the parameter domain retained in this work.
Black solid curves mark $\epsilon_3=0$, and hatching marks the portions of the JP regions below the Kerr reference band.
Right: enlargement of the boxed region for $\omega/\omega_K=1$.
Dash-dotted curves show constant $a$ and constant $\epsilon_3$, with their values labeled.
}
\label{fig:kerr_band}
\end{figure*}

\subsection{Non-Kerr signatures and beyond Keplerian motion}
\label{subsec:kerr_band}

The negative-$\epsilon_3$ branch in Fig.~\ref{fig:parameter_flow} lies below the trajectory generated by varying the orbital radius in the Schwarzschild spacetime, whereas increasing the prograde spin shifts the observables toward larger intercepts. This contrast suggests a way to identify departures from Kerr through the joint values of $k$ and $b$. We now test whether this distinction survives departures from Keplerian motion by asking whether a Kerr model with a different spin and orbital angular velocity can reproduce the same pair $(k,b)$.

For this comparison, we fix $r_s=8M$ and $\theta_o=17^\circ$ and retain the emission model of Sec.~II. The Keplerian angular velocity $\omega_K$ is evaluated separately in each spacetime using Eq.~(12), and $\omega/\omega_K$ is varied independently of the spin and deformation. Ratios below, equal to, and above unity describe sub-Keplerian, Keplerian, and super-Keplerian motion, respectively. We construct the Kerr reference region by varying $a$ from $0$ to $1$ and allowing $0.5\leq\omega/\omega_K\leq1.5$. Its predictions occupy the gray band in Fig.~3, with lower and upper dashed boundaries corresponding to $a=0$ and $a=1$. The band thus incorporates freedom in both the Kerr spin and the orbital angular velocity.

The colored regions show the JP predictions for $\omega/\omega_K=0.7$, $1.0$, and $1.3$. Color denotes $\epsilon_3\in[-10,10]$, and the black curves mark the Kerr limit $\epsilon_3=0$. The hatched portions fall below the Kerr band and correspond to low-spin configurations with negative $\epsilon_3$. At the same slope, their intercepts are smaller than those attained anywhere in the Kerr reference band. No choice of the Kerr spin and orbital angular velocity can therefore reproduce these points. Such regions appear at all three displayed ratios, demonstrating that distinguishable non-Kerr signatures persist whether the motion is slower or faster than the Keplerian rate. The Kerr band thus provides a direct consistency test. Points inside the band admit a Kerr interpretation, while points below its lower edge have intercepts smaller than any Kerr model in the reference band can produce at the same slope. The latter identify departures from Kerr without requiring prior knowledge of the orbital angular velocity.

Increasing $\omega/\omega_K$ shifts the colored regions toward larger $k$ and smaller $b$. The dependence of the slope $k$ on source motion is expected, whereas the dependence of the intercept $b$ on $\omega$ can be understood from the different propagation times of the image orders. At a fixed observing time, the higher-order image corresponds to an earlier emission time than the direct image. Changing the orbital angular velocity therefore changes the orbital-phase separation between the source positions represented by these images, and hence their angular separation and the ridge intercept.

The enlarged right panel also illustrates how the ridge observables can be used to constrain spin and deformation together. At $\omega/\omega_K=1$, the curves of constant $a$ and constant $\epsilon_3$ show that increasing the spin mainly raises the intercept, whereas changing the deformation also shifts the slope appreciably. These different responses allow measurements of $k$ and $b$ to jointly constrain $a$ and $\epsilon_3$. An independent measurement of $\omega$ can likewise supply the orbital information required for this analysis: the predictions are then evaluated at the measured angular velocity, with the corresponding ratio $\omega/\omega_K$ determined by each spacetime.

\section{Summary and discussion}
\label{sec:discussion}

We investigated the ability of black hole flare autocorrelation to distinguish departures from the Kerr geometry, using the parametrized Johannsen-Psaltis spacetime with a single metric deformation parameter as a test case. Numerical ray tracing of a compact hot spot on an equatorial circular orbit produces time-angle autocorrelation maps with lensing-induced secondary ridges. The slope and intercept of a secondary ridge provide a pair of observables that encode information about the spacetime.  The results show that the spin and metric deformation parameter affect the ridge observables differently. Increasing the prograde spin primarily raises the intercept, whereas negative $\epsilon_3$ shifts the observables toward larger slopes and smaller intercepts. With the orbital radius, inclination, and $\omega/\omega_K$ specified, the different responses of the slope and intercept allow joint constraints on the spin and metric deformation parameter.

For $r_s=8M$ and $\theta_o=17^\circ$, we constructed a Kerr reference band spanning $0\leq a\leq1$ and $0.5\leq\omega/\omega_K\leq1.5$. Low-spin configurations with negative $\epsilon_3$ extend below its lower boundary, with intercepts smaller than the Kerr minimum at the same slope. Their autocorrelation signatures cannot be reproduced by changing the Kerr spin and orbital angular velocity. This separation occurs at each of the three investigated ratios, $\omega/\omega_K=0.7$, $1.0$, and $1.3$, and therefore persists when the source moves more slowly or more rapidly than the Keplerian rate.

For the low-spin configurations identified above, the pair of ridge observables distinguishes the effects of a negative JP deformation from those of changes in the Kerr spin and orbital angular velocity within the reference region. A natural next step is to extend the analysis to other parametrized metrics and black hole solutions in specific theories of gravity. Comparing the resulting autocorrelation signatures would reveal which features are shared by different non-Kerr metrics and which depend on the particular spacetime, clarifying the sensitivity of the autocorrelation to different departures from Kerr.

On the astrophysical side, the present analysis is restricted to equatorial circular motions, leaving non-circular and non-equatorial trajectories unexplored \cite{Aimar:2023kzj,Huang:2024wpj}. Moreover, the compact hot-spot model does not account for larger-scale emission structures \cite{Emami:2022ydq} or turbulence-driven intrinsic variability \cite{Dexter:2010pe,Lee:2020pvs,Nathanail:2020wap}. Future work can extend the analysis to more general trajectories and use numerical simulations to model spatially extended, evolving emission \cite{Jiang:2024gtk}, allowing us to investigate how these effects alter the autocorrelation signatures of departures from Kerr. 

\section*{acknowledgements}
We thank Yehui Hou and Xi Wan for helpful discussions. The work is partly supported by  National Natural Science Foundation of China
(NSFC) Grants No. 12275004, 12547132, 12547127, and No. 12588101. Zhenyu Zhang is also supported by the Postdoctoral Science Preferential Funding of Zhejiang Province under Grant No. ZJ2026086.

\bibliographystyle{utphys}

\bibliography{JPhspt}

\clearpage

\appendix

\onecolumngrid

\section{Hot-spot emission and radiative transfer}
\label{app:hotspot_emission}

The center of the hot spot follows the equatorial circular orbit
\begin{align}
	x_s^\mu(t)
	=
	\left(
	t,r_s,\frac{\pi}{2},\phi_s(t)
	\right),
	\qquad
	\phi_s(t)=\phi_{s,0}+\omega t.
	\label{eq:hotspot_trajectory}
\end{align}
Its four-velocity is
\begin{align}
	u_s^\mu
	={}&
	u_s^t(1,0,0,\omega),
	\nonumber\\
	u_s^t
	={}&
	\left(
	-g_{tt}-2\omega g_{t\phi}
	-\omega^2g_{\phi\phi}
	\right)^{-1/2},
	\label{eq:hotspot_four_velocity}
\end{align}
where the metric components are evaluated at
$r=r_s$ and $\theta=\pi/2$.

For an emitting point
$x_p^\mu=(t_p,r_p,\theta_p,\phi_p)$, we define its spatial separation from the
instantaneous hot-spot center by
\begin{align}
	x^2={}&
	\left[
	r_p\sin\theta_p\cos\phi_p
	-r_s\cos\phi_s(t_p)
	\right]^2
	\nonumber\\
	&+
	\left[
	r_p\sin\theta_p\sin\phi_p
	-r_s\sin\phi_s(t_p)
	\right]^2
	+r_p^2\cos^2\theta_p
	\nonumber\\
	={}&
	r_p^2+r_s^2
	-2r_pr_s\sin\theta_p
	\cos\!\left[\phi_p-\phi_s(t_p)\right].
	\label{eq:hotspot_spatial_distance}
\end{align}
The corresponding emissivity profile is
\begin{align}
	J(x)
	=
	J_0\exp\left(-\frac{x^2}{2s^2}\right),
	\label{eq:hotspot_emissivity_appendix}
\end{align}
where $J_0$ is an overall normalization and $s$ determines the hot-spot size.
We adopt $s=0.3M$ throughout the numerical calculations. Since the orbital
phase in Eq.~\eqref{eq:hotspot_spatial_distance} is evaluated at the emission
time $t_p$, the emissivity varies during photon propagation.

We assume that the hot spot is optically thin and neglect absorption. The
invariant radiative-transfer equation along a null geodesic is then
\begin{align}
	\frac{\mathrm d}{\mathrm d\lambda}
	\left(\frac{I_\nu}{\nu^3}\right)
	=
	\frac{j_\nu}{\nu^2},
	\label{eq:invariant_radiative_transfer}
\end{align}
where $I_\nu$ and $j_\nu$ are the specific intensity and emissivity,
respectively, $\nu$ is the photon frequency measured in the local rest frame,
and $\lambda$ is an affine parameter. We further assume a frequency-independent
emissivity, $j_\nu=J$. Introducing the redshift factor
\begin{align}
	g
	\equiv
	\frac{\nu_o}{\nu_e}
	=
	\frac{-k_\mu u_o^\mu}
	{-k_\nu u_s^\nu},
	\label{eq:hotspot_redshift_factor}
\end{align}
and integrating Eq.~\eqref{eq:invariant_radiative_transfer}, we obtain
\begin{align}
	I_o
	=
	\nu_o\int g^2J\,\mathrm d\lambda
	\propto
	\int g^2J\,\mathrm d\lambda.
	\label{eq:observed_intensity_appendix}
\end{align}
The integration is performed along null geodesics traced backward from the
observer's screen. At every point along a ray, the hot-spot position is
evaluated at the corresponding emission time $t_p$. This slow-light
prescription accounts for the source evolution during photon propagation and,
in particular, for the different emission phases associated with the direct
and higher-order images.

\end{document}